\documentclass[11pt,a4paper]{article}
\usepackage[left=1in, right=1in, top=1in, bottom=1in]{geometry}

\usepackage{authblk}
\usepackage{enumerate}
\usepackage{tabularx}
\usepackage[dvipsnames]{xcolor}
\usepackage{hyperref}
\usepackage{multirow}

\usepackage{amsmath,amssymb,amsfonts,mathtools}
\usepackage[mathscr]{euscript}
\usepackage{graphicx}
\graphicspath{ {./images/} }

\definecolor{red}{cmyk}{0,1,0.77,0.3}
\definecolor{yellow}{cmyk}{0,0.05,0.89,0.03}
\definecolor{green}{cmyk}{0.65,0,0.92,0.15}
\definecolor{blue}{cmyk}{0.9,0.3,0.2,0.07}
\definecolor{grey}{cmyk}{0,0,0,0.5}

\newcommand*{\email}[1]{\href{mailto:#1}{\url{#1}} }

\begin{document}

\title{A Non-causal Reconceptualization of Quantum Field Theory}
\date{\vspace{-5ex}}

\author[1]{\small
Christopher Thron
}
\affil[1]
{\small Department of Science and Mathematics, Texas A\&M University-Central Texas\\
\email{thron@tamuct.edu}}

\maketitle

\noindent{\bf Abstract:} 

Quantum field theory currently has a single standard mathematical characterization (the Standard Model), but no single accepted conceptual framework to interpret the mathematics. Many of these conceptualizations rely on intuitive concepts carried over from classical physics (such as “particle” and "causality").
In this paper, instead of relying on classical concepts we attempt to infer a conceptualization directly from the mathematics. This reconceptualization leads to a physical reinterpretation of the paths involved in the Standard Model’s action integral as traces of an accumulative process that occurs within an ``extraverse'', of which our observable spacetime is a single ``slice''. We briefly outline a rigorous mathematical model which validates the physical reinterpretation, and leads to predictions that are potentially verifiable by experiment. We contrast our model  with other popular interpretations of quantum mechanics. We describe some of the metaphysical consequences of our proposed perspective. Finally we present some speculations about an alternative approach that could possibly lead to a more satisfying, intuitively-graspable theory.

\noindent{\bf Keywords:} quantum field theory; causality; process; spacetime; universe; many-worlds; free will.\\


\maketitle

\section{Introduction}

The gist of this section may be summed up in the following slogans:
\begin{enumerate}
\item
Physics is simple, but observations may be complicated.
    \item 
Physics walks on two legs: analogy and mathematics;
\end{enumerate}

\subsection{Physics, observations, and complications }

Complications in physics are often due to inferences drawn from peculiarities arising from observational viewpoint. For example, living in a world beset by friction  
led naturally to Aristotelian physics, in which objects remain at rest in their ``natural state'' unless they are pushed. Humankind's  geocentric observational viewpoint motivated Ptolemaic astronomy. Everyone (physicists included) up to the 20\textsuperscript{th} century took as fact that time was absolute, the same for every observer. These conclusions were so natural and ``self-evident'' that they were scarcely questioned in their day.

All of these theories were eventually supplanted by new alternatives that provided more accurate predictions with simpler mathematics. These new theories were not abstruse or complicated, but did require a radical change of thinking. Newton's Law of uniform motion is readily grasped, once one accepts the fact that the "slowing down" of objects is produced by secondary causes. The heliocentric view is standard fare for primary-school science classes. Even special relativity is not hard to understand, once one gets accustomed to the idea that changes between reference frames produce a sort of ``rotation'' that mixes space and time.  

 The currently-accepted fundamental theory of particle physics (the so-called Standard Model) is beset with extreme mathematical complications. Perhaps this complication is a sign that some fundamental presumptions should be questioned. In this paper, we consider two implicit notions that underlie conventional particle physics---namely causality and the notion of ``particle''.  We suggest that these notions are only muddying the waters, and a clearer and simpler view emerges if they are thrown out.
 


\subsection{Analogy and mathematics in physics}
The analogical aspect of physics consists of conceptual or mechanistic models, paradigms, and examples that show comparable behavior to the physical system under consideration. 
The importance of analogy in physics is abundantly demonstrated throughout history.  Aristotle had his crystal spheres; Newton envisioned particles of light, while Huygens likened light to water waves. Maxwell built on this idea, and his terminology (e.g. ‘electromagnetic stress tensor’) showed that he conceived of light as a wave-like disturbance of a medium. The statistical mechanics of gases is based on the conceptual picture of gases as ensembles of tiny billiard balls bouncing off each other. This analogy was taken further in Einstein’s theory of Brownian motion. The Thomson “plum pudding” model  of the atom gave way to Rutherford’s planetary visualization, which set the stage for Bohr’s idea of electrons as de Broglie waves wrapping around the nucleus. Special relativity pictures light rays as absolute yardsticks for measuring space and time; and general relativity slightly modified the picture so they became geodesics in an intrinsically curved, four-dimensional space.

The mathematical aspect of physics involves mathematical formulations, equations, and theorems.
Mathematics makes analogies precise---and as the mathematical formulations are progressively simplified they predict new physics, and suggest further conceptual analogies. Thus Planck’s Law for black-body radiation, which was an empirical mathematical expression, led to Einstein’s concept of quanta.  This in turn led to de Broglie’s momentum-wavelength relation, which provided the dispersion relation that motivated Schr\"{o}dinger’s equation.

The above examples clearly demonstrate the fruitful and longstanding dialog between analogy and mathematics in physics. But in mainstream quantum mechanics, the dialog is cut off. The predominant Copenhagen interpretation shrugs its shoulders and says, “Analogies no longer apply. From here on out we rely only on mathematics.”  Since that time, quantum mechanics (and its offspring quantum field theory) have mainly relied on mathematical symmetries and elegance for theoretical insights. So quantum theory continues to hop along on one leg. To be sure, there have certainly been analogies between quantum field theory and other physics (notably condensed matter physics). However these have mostly involved mathematical techniques that can be applied to both (e.g. renormalization group methods), while remaining agnostic to the underlying ``mechanisms''.

Are analogies really necessary for continued progress in physics? History shows that analogies have served physics amazingly well---but invariably they have turned out to be inadequate, as the examples above show. Typically, after initial successes the paradigm entailed by the analogy becomes less and less adequate, and requires more and more complicated modifications (e.g. `epicycles') to accommodate observations, When this happens, it is time to look for a different paradigm.
Replacing one paradigm with another may involve throwing out the most basic concepts and descriptors in favor of others. For example, “force” as the centerpiece of Newton’s mechanics, in modern physics plays a secondary role as the gradient of energy, which is considered as the more fundamental concept. Similarly the “ether” which was foundational to Maxwell’s understanding of electromagnetism has been supplanted by the ether-less  spacetime manifold. 

Nonetheless, the eventual obsolescence of conceptual models/analogies in no way detracts from their temporary necessity.  Indeed, these analogies are what give us the feeling that we really “get” something, by relating unfamiliar phenomena to familiar situations that we have an intuitive "feel" for. It is interesting that the same sort of process takes place in deep neural networks, which learn to recognize images by identifying increasingly complex features that are built on previously-learned features\cite{lecun2015deep}.
One effective particularly technique within deep learning is “transfer learning”, in which a neural network that has been trained on one data set is then exposed to a different type of data \cite{bozinovski2020reminder}. The higher layers of the network are then allowed to adapt to the new data, while the lower layers (representing the basic features learned from the previous dataset) are held fixed.

There is a tendency in modern science to emphasize the role of mathematics in physical discoveries.
Certainly the most spectacular modern theories are tour de forces of mathematical gymnastics: the Standard Model,  black hole physics, string theory \ldots the list goes on.  From these examples We derive the impression that new physics falls out of the mathematics. for example, in a public lecture on ``The Beauty of Calculus'', the prominent mathematician and math popularizer Steven Strogatz describes Maxwell's discovery of his equations as follows:
\begin{quotation}
Maxwell looked at \ldots his equations.  As he starts manipulating them \ldots and he doesn't know what he's looking for. He just knows that he wants something to come out,…
 and at some point he recognizes a new equation that has come out
because it's the same equation that describes the spread of ripples on a pond. It's the equation for how waves move,
and except that this is a wave of electricity and magnetism. \ldots He plugs in the numbers and it turns out it propagates at the speed of light, which had been measured around
that time for the first time.  \cite{Strogatz2019}
\end{quotation}
This makes for a nice story, but that’s not at all what happened. Maxwell was certainly familiar with the experiments of Thomas Young that demonstrated the wave nature of light. Furthermore, Weber had previously shown (in 1856) that the speed of light had something to do with electromagnetism.  So it’s simply not true that Maxwell ``didn’t know what he was looking for’’.

Maxwell’s reasoning is very clearly delineated in his 1861 paper. His starting point was not the equations, but an analogy between lines of force in a solid medium and those produced by charges and magnets. Maxwell postulated that these lines of force represented states of tension and compression in a physical ether. He concluded that just as waves propagate in solid media, so there should be waves that propagate through the ether. He was then able to characterize these waves (and their speed) using the equations he had assembled from Gauss, Faraday, and Ampere. So unlike Strogatz’ presentation, the mathematics did not motivate Maxwell’s discovery, but rather played a technical role in enabling a precise characterization of the waves that Maxwell already expected to occur.

Returning to the  situation with modern quantum physics, while the mathematics has made great strides the analogical side has dragged along like a game leg. Various “interpretations” or “explanations” of quantum theory have been advanced. Some of these are quite popular (for example, the ``many-worlds'' interpretation) but none has gained anything approaching majority consensus within the physics community.  These interpretations are commonly viewed as  attempts to show what is “really” going on “behind the scenes”, as it were. But it is better to think of such interpretations as temporary vehicles, as all previous analogies in physics have been. 

Even if we claim to free ourselves from analogies, they nonetheless exert unconscious influence on the way we think about the physics. This comes out in the language we use to describe what the equations represent: for example, we say that the quantum mechanical wavefunction describes the probability distribution of a `particle' (or `particles') which are the `causes' measurement outcomes.  
The drought of fresh insights on the analogical front indicates that perhaps some of our tacit assumptions should be critically re-examined. 
In the next section, we present two experiments that point to fundamental 
difficulties with the common notions of `particle' and `cause-and-effect'.

\section{Experimental evidence against common notions}\label{sec:quantumGame}

The following experiments are not new, but they are usually subject to different interpretations (e.g. in many-worlds, pilot wave or transactional theories) or left uninterpreted (Copenhagen). We argue that on the contrary they point in a different direction, away from `` particles'' and ``causes''.

\subsection{The Aspect experiment}

The 2022 Nobel prize was awarded to John Clauser and Alain Aspect (shared with Anton Zeilinger) for experiments conducted in the 1970's and 1980's \cite{clauser1969proposed,clauser1978bell,aspect1981experimental,aspect1982experimental}. 
These experiments (which we will refer to generically as the ``Aspect experiment'')   established the `weird' phenomenon of quantum entanglement.   But as we shall see, the result is only weird if we presume that  spacetime events are produced by objects (quanta) that travel through spacetime.

\begin{figure}[h]
    \begin{center}
        \includegraphics[scale=0.5]{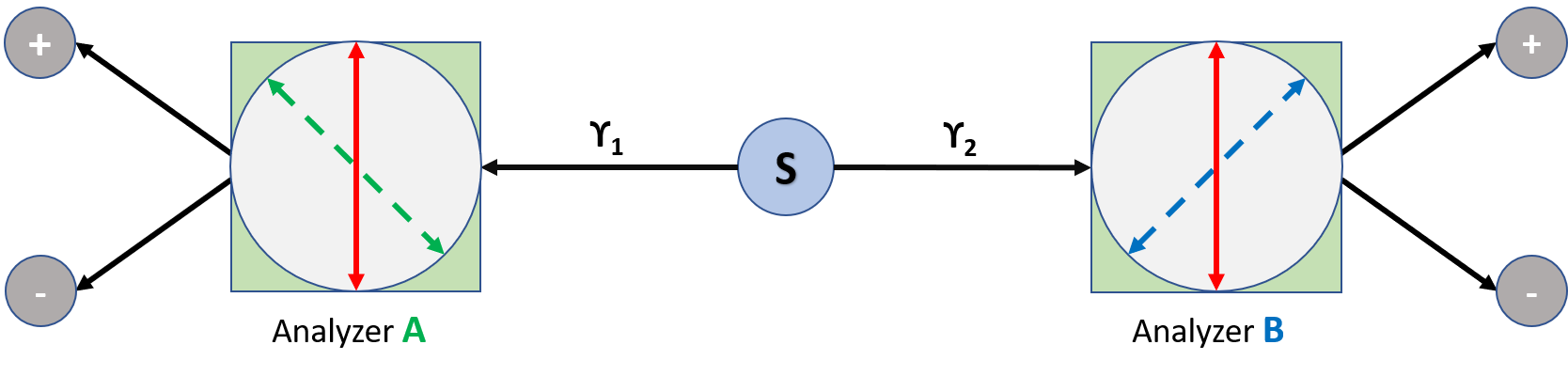}
    \end{center}
       \caption{Schematic representation of Aspect's experiment, as described in \cite{aspect1982experimental}. The red line indicates the initial orientation of $A$ and $B$. The green dashed line indicates $A$ is oriented in the counter-clockwise direction while the blue dashed line indicates $B$ is oriented in the clockwise direction. The angle between $A$ and $B$ is used to calculate the probability of the outcome.\label{fig:Aspect} }
\end{figure}

Figure~\ref{fig:Aspect} is a schematic representation of the Aspect experiment. The figure shows a source of circularly polarized photon pairs, which fly off in opposite directions and are detected by the two detectors (left and right). Each detector can be set at two different angles. Any electron detection can produce one of two outcomes:  parallel or perpendicular. 

There are four different ways to align the two detectors:  left=green, right=red;  left=red, right=blue;  left=red, right=red; and left = green, right = blue.  For each different alignment, there are four different possible outcomes. For simplicity we will characterize the outcomes as ``same'' (both parallel or both perpendicular) or ``different'' (one parallel and one perpendicular).
 From thousands of repeated measurements we obtain the following results:
 \begin{itemize}
 \item
 When left=green and right=red, the measurements are the same 85\% of the time; 
 \item
 When left=red and right=blue, the measurements are the same 85\% of the time:
 \item
 When left = red and right = red, the measurements are the same 100\% of the time.
 \end{itemize}
 Since both green and blue differ from red at most 15\% of the time, it follows from  elementary probability theory (inclusion-exclusion) that all three detectors should agree at least $100-15-15 = 70$ percent of the time.  However, when measurements are performed with the green-blue configuration, it turns out that the detectors agree only $50$\% of the time. The same result holds even if the detectors' configurations are changed ``randomly`` after the photon pair is produced. This would seem to be impossible, unless the detectors are in collusion and know each others' settings.\footnote{For a more extensive discussion of the collusive aspect of the experiment, see \cite{welsch2021thequantum}.}  

\begin{figure}[h]
    \begin{center}
        \includegraphics[scale=0.5]{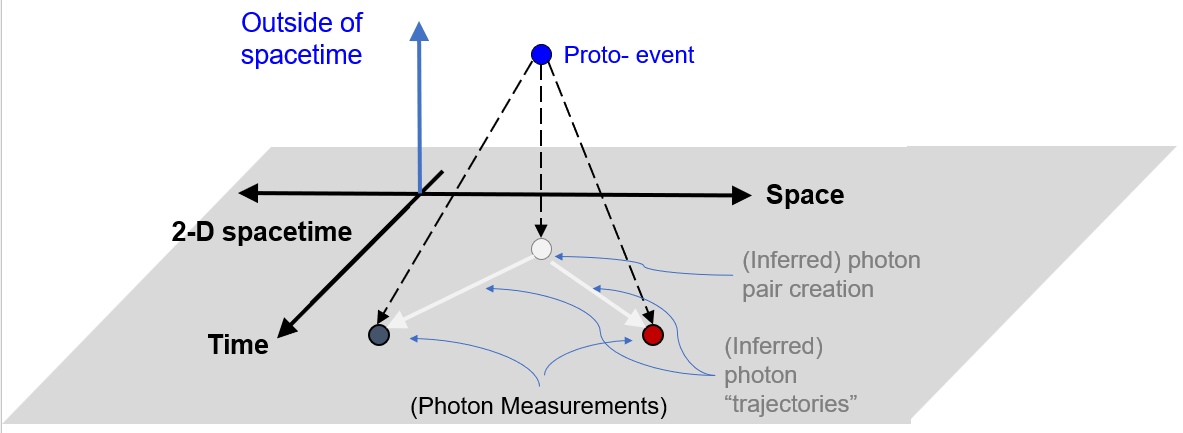}
    \end{center}
       \caption{Illustration of non-causality. Something happens ``out there'' outside of spacetime, producing the two detections.  The existence of a prior (in time) photon decay event and subsequent trajectories is seen as an artifact.}\label{fig:artifact}
\end{figure}

Usually the Aspect experiment is cited as demonstrating ``spooky action at a distance'---i.e. if one detector is changed at the last minute, then the outcome at the other detector is instantaneously affected.  But from an alternative viewpoint, we may say instead that it shows there is no such thing as a `passive' measurement that unobtrusively and impartially records `objective' outcomes. Instead, any measurements of an interaction must necessarily be considered as intrinsic to the interaction.
To see how this might work, consider a simplified ``universe'' which has only one space and one time dimension.  Then  spacetime can be represented as a 2 dimensional plane, where the two axes represent space and time respectively (see Figure~\ref{fig:artifact}). We now  introduce into the picture an additional dimension, which we'll call the \emph{process dimension}, which is \emph{outside} of space and time. We suppose that ``proto-events'' occur in this higher-dimensional ``extraverse'', and the influence of such proto-events propagates down along the process dimension. Figure~\ref{fig:artifact}  shows how this influence results in correlated measurement outcomes while producing the illusion of a strict causal relationship between  events and  detections within spacetime. In this scenario,  nothing actually ``travels'' between the creation and detection of the photons---there's only the illusion of movement, which comes from the fact that both the creation and detection originate from a common source.  
  
  The picture in Figure~\ref{fig:artifact}  defies the near-sacred presumption of causality in physics. But as we saw in the introduction, sacred idols in physics have been toppled before. After all, what's the evidence for causality? Only that it's an ``obvious fact'' from everyday experience. But this very obviousness should set off a mental smoke alarm---for  all too often``obvious facts'' have led physicists badly astray. Indeed, despite thousands of years of speculation,  scientists and philosophers have failed to make a rigorous distinction between causality and correlation \cite{pittphilsci11913}. This failure should deepen our suspicions that  causality may be nothing but smoke and mirrors. 

\subsection{The Elitzer-Vaidman quantum bomb-tester}

The ``quantum bomb tester'' refers to a thought experiment designed to illustrate the phenomenon of ``interaction-free measurement''. Elitzur and Vaidman specified a procedure whereby one could test whether a bomb is a dud or not without even looking in the direction off the bomb. Subsequent experiments (without bombs) showed that interaction-free measurements of this type are indeed possible.

The original article  \cite{elitzur1993quantum} contains a very readable description of the thought experiment. An alternate presentation is given in \cite{robens2017atomic}, which also  describes an experimental test of the effect.  See \cite{yang2023interaction} for an experimentally-realized practical application of interaction-free measurement.

\begin{figure}[h]
    \begin{center}
        \includegraphics[scale=0.5]{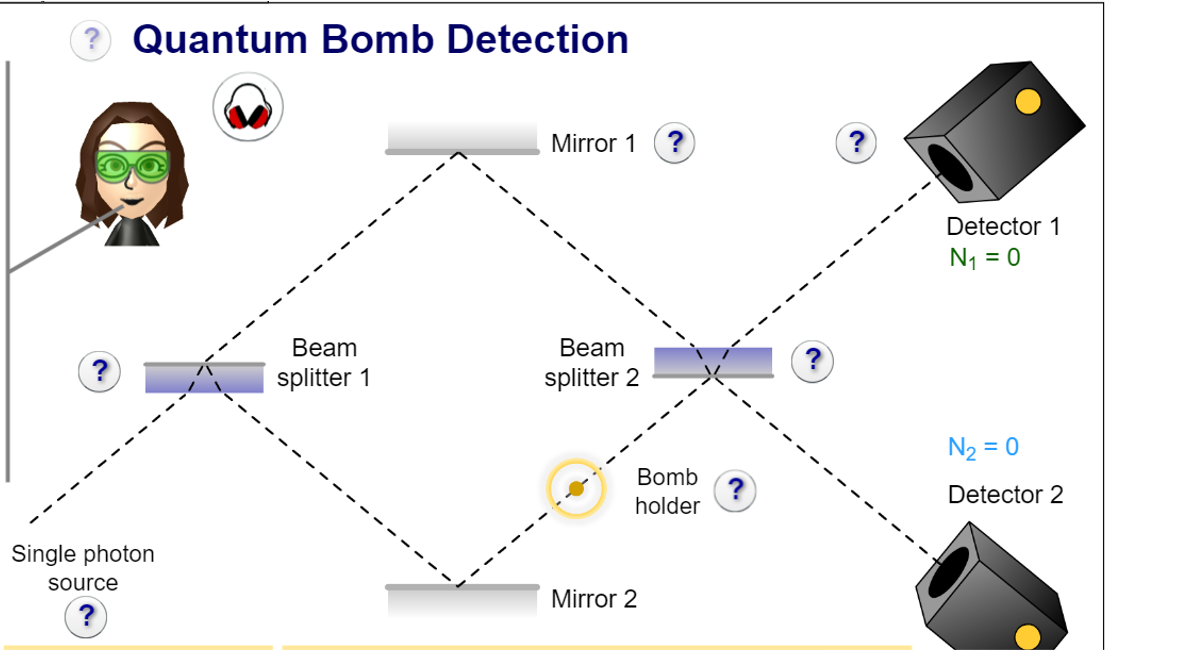}
    \end{center}
       \caption{Elitzur-Vaidman quantum bomb detector (from \url{https://www.st-andrews.ac.uk/physics/quvis/simulations_html5/sims/})}\label{fig:qdd}
\end{figure}

Figure~\ref{fig:qdd} shows the basic experimental setup.  A  source that produces isolated photons is located a lower left. The photon impinges on various beam splitters as shown, so that multiple paths are possible. The apparatus is configured so that  no photon can reach Detector 1 if the path via Mirror 2 is open, due to destructive interference. 
The lower path is not open if and only if a photon passing through interacts with the bomb, causing it to explode
If the lower path is not open, then quantum mechanics predicts three possible outcomes:
either the bomb explodes (with 50\% probability); or the photon is detected at Detector 2 (25\% probability); or the photon is detected at Detector 1 (25\% probability).
In the latter case, the live bomb was detected without exploding it, since a detection at Detector 1 is impossible if the bomb is unarmed.
While 25\% recall is remarkable enough, Hosten et al.   have shown that the setup can be modified so that live bomb detection occurs before detonation with arbitrarily high probability, by increasing the number of possible trajectories \cite{hosten2006counterfactual}.

\begin{figure}[h]
    \begin{center}
        \includegraphics[scale=0.5]{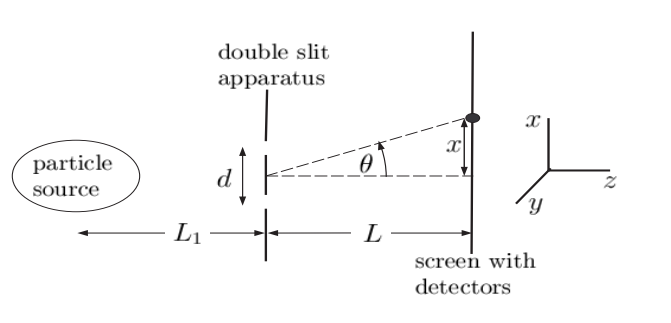}
    \end{center}
       \caption{Generic double slit experiment (from \cite{rolleighdouble2010}).}\label{fig:qdd2}
\end{figure}

A similar effect (though much less striking) can be observed in  two-slit experiments  that demonstrate the wavelike diffraction of particles such as photons or electrons. Such experiments are described in many textbooks as well as popular presentations of quantum mechanics (apparently Richard Feynman started this practice, see \cite{rolleighdouble2010}). The setup is shown in Figure~\ref{fig:qdd2}. If both slits are open, then the interference pattern has periodic nulls---but if only one is open, the interference pattern is unimodal with no nulls. Suppose we have a screen with two slits, but we  don't know whether or not the second slit is blocked (in a bomb detection scenario, the   second slit is blocked if and only if the bomb is activated). The source emits particles one at a time. If a particle detection is made at a null location, then we know immediately that the second slit is closed---so apparently we have determined the open/closed status of the second slit by passing a particle through the first slit. 
It boggles the imagination to consider how a particle that passes one way can tell us anything at all about a route that it has not taken. But the result is perfectly compatible with the scenario  in Figure~\ref{fig:artifact}, because there are no ``traveling particles'', only detections whose statistics are determined by the entire experimental setup. 

In the next section we develop an analogy inspired by wireless communications technology that includes the same characteristics as Figure~\ref{fig:artifact}.  We find that not only does the analogy explain both spooky action at a distance and interaction-free measurement, it also accounts for the Born rule that quantum measurement probabilities are given by squared amplitudes of complex wavefunctions.

\section{Overview of the extraverse model}

We have suggested that new analogies are required to explain puzzling behavior which does not fit our settled notions of how particles should behave.  We now have an advantage over previous generations in that we have a rich new source from which to draw  examples, namely modern technology. Indeed, quantum particles are carriers of information, and the universe may be conceived of as an information communications system. This suggests that modern communications technology might give us some insight into  particle physics. 

Wireless digital communication does indeed have a number of apparent similarities with quantum particle dynamics. Both involve the transmission of  discrete information using waves, Both are described mathematically with a complex field defined at all points in spacetime.
Furthermore, there are scenarios in which information detection probabilities are given by the absolute squares of complex field amplitudes.
This happens in particular in incoherent signal detection, when the information-carrying signal is modulated by noise.

The mathematical similarity between quantum and signal-processing probabilities is brought out especially strongly in the  path integral formulation of quantum mechanics, due to St\"uckleberg and Feynman. The path integral representation for the transition probability to a state $u$ is written as:
\begin{equation}\label{eq:prob}
P(u) \propto \left| \sum_{ \gamma \in \Gamma_u} e^{iS(\gamma)} \right|^2,
\end{equation} 
where $\Gamma_u$ is a space of paths corresponding to the state $u$, and $S(\gamma)$ is the action (path integral) associated with the path $\gamma$ (we have written this as a sum instead of the usual integral for conceptual simplicity). To a wireless communications engineer, the sum inside the absolute values in \eqref{eq:prob} is immediately recognizable as the complex representation of the signal received by  a wireless receiver (such as a mobile phone) in an environment with obstacles that produce multiple signal paths (``multipaths'') between signal source and receiver. Furthermore, the squared amplitude of this complex signal corresponds to the signal power.

\begin{figure}[h]
\begin{center}
\includegraphics[width=5.in]{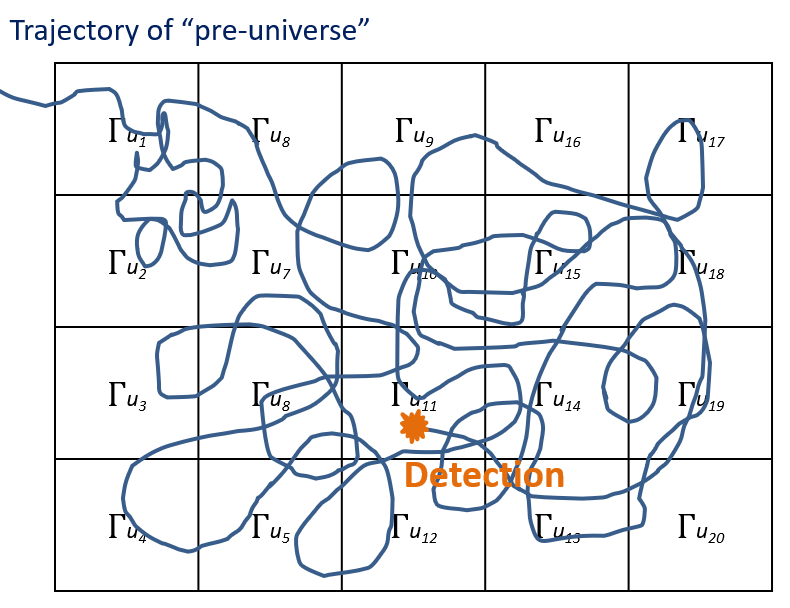}
\end{center}
\caption{ Stochastic process in which a system wanders through a space of possible system states, until an ongoing accumulative process attains a certain threshold, and the system is actualized. (The ``system''  can represent a wireless receiver in a communications scenario, or a physical system in the quantum scenario.) \label{fig:process}}
\end{figure}

Figure~\ref{fig:process} schematically represents a wireless communications scenario in which a ``receiver'' wanders through the space of possible locations. The path integral corresponds to an ongoing signal accumulation, such that the signal is detected when the accumulation reaches a certain threshold. To complete the analogy with quantum mechanics, we replace the receiver with a ``status indicator'' that wanders among possible system configurations during the process. A similar path-integral accumulation occurs during the wandering process, until a threshold is reached. When the threshold is achieved, the configuration in which the status indicator resides determines the observed physical configuration.  Note that the ``physical configuration'' we are referring has extent in both space and time: for example, it would include both the emission and detection of particles in an experiment. For a complete physical picture, the configurations in the configuration space should correspond to states of the entire universe.

By imposing a few simple conditions on  our proposed stochastic process, we may recover the state probabilities \eqref{eq:prob}. A complete mathematical derivation may be found in \cite{thron2015b}): this paper also explains physical consequences of the theory that differ from conventional quantum mechanics, which could conceivably be used to test the theory.

Although we may apply this analogy to individual physical systems, for a complete account of physical experience it is necessary to take the system as the entire spacetime universe.
In this case, we may envision this process as  an ``extraverse'' that evolves in some unobservable, non-spacetime dimension until the accumulated amplitude achieves a certain level,  at which point the  configuration matches ``our'' universe. As a visual analogy, consider a giant soap bubble that  contorts as it floats through the air (for a marvellous video of this, see \cite{Day}). Over time the water in the membrane evaporates, until the remaining water content is no longer sufficient to maintain the necessary surface tension \cite{Troutner}. Eventually the membrane fails at a single point, and the bubble disintegrates as though rolled up. Although it appears as if receding edge of the vanishing bubble is a traveling object  (just as we  observe ``objects'' that travel through space and time), in fact the moving edge simply traces over the shape of the membrane that  was already there. (Naturally there are limits to this analogy---the extra-universal process by which the universe is \emph{not} a process in time, and we are not necessarily implying that the universe is ``rolling up'' as time progresses.) 

\begin{figure}[h]
\begin{center}
\includegraphics[width=5.in]{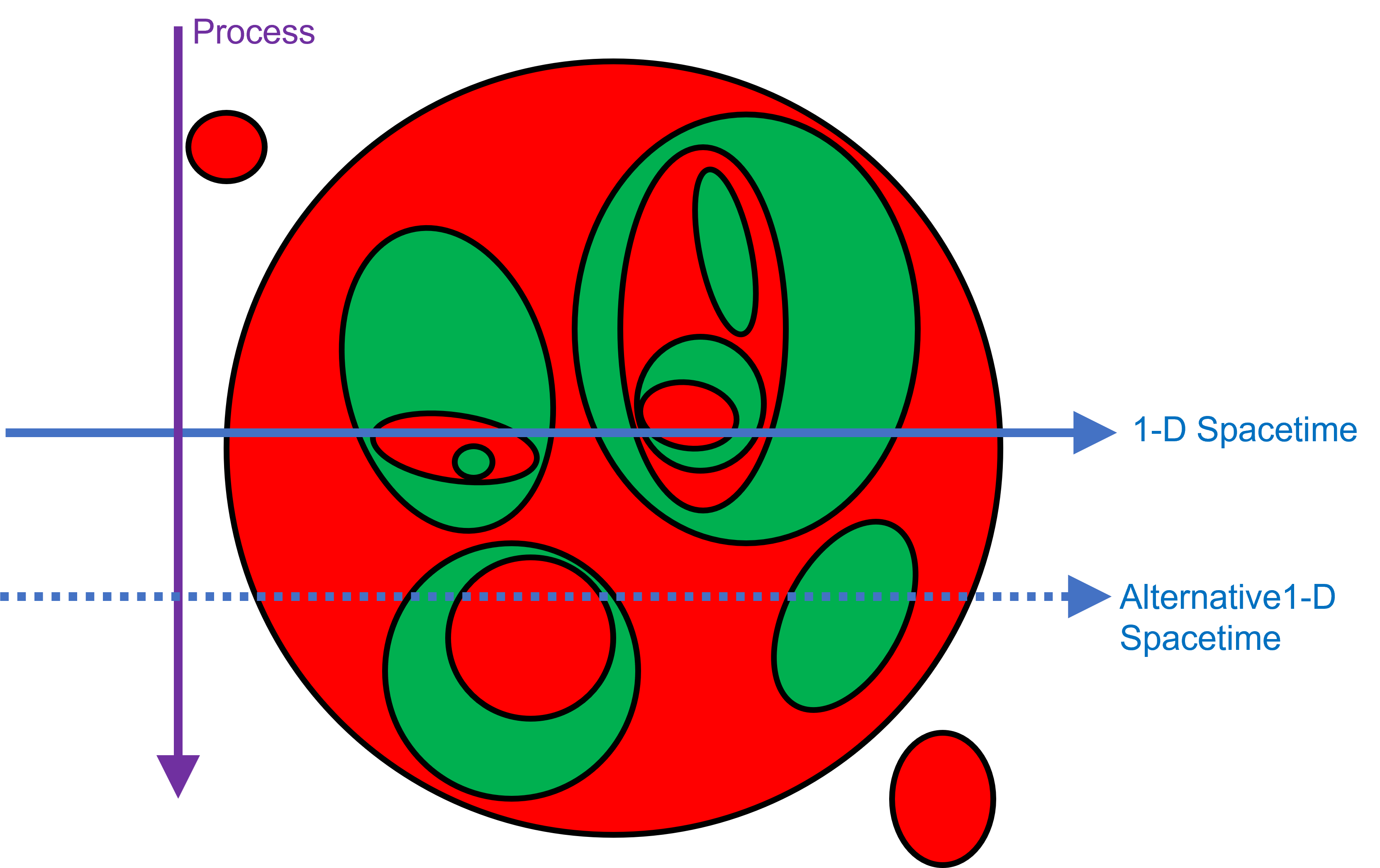}
\end{center}
\caption{ Evolution of the ``extraverse'' as a process unfolding in a dimension outside of spacetime, which produces spacetime when the process attains a certain threshold. \label{fig:0d}}
\end{figure}

A simple schematic of this intuitive picture is given in Figure~\ref{fig:0d}. For ease of representation we show a spacetime universe with only one dimension (which could be either a single point that advances in time, or a timeless spatial expanse) which lies within a larger ``extraverse''. The red and green bubbles in the picture evolve along the process dimension (downward arrow), and the spacetime universe is a single slice or section of this larger picture. Other slices are also possible, and would correspond to alternative universes.   

A possible objection to this proposal is that since it treats our observed spacetime universe as an outcome and not process, it does not explain why we experience time unfolding progressively. But neither does conventional physics. Conventional physics simply presumes that the equations governing phenomena have time as a dependent variable. The same equations can be derived (as approximations) under our proposal. The difference is that the extraverse model explains our experience of unfolding time as an artifact of how we (as phenomena within spacetime) relate to other phenomena within spacetime. Our experience is related to the information that we are able to assimilate and act upon---and we are able to access information that flows from the causal past within our spacetime because of peculiarities in our own construction.

Note that the extraverse presented here is completely different from the multiverse which is associated with the theory of cosmic inflation \cite{linde2017brief,davies2004multiverse}.  The multiverse includes disjoint universes that are spun off as a result of a temporal process and subsequently evolve independently in time (with perhaps some interactions). In the extraverse picture, possible universes are ``slices'' do not evolve in parallel but rather blend into each other through a continuous process of distortion.

\section{Conceptual implications of the extraverse model}

We've proposed that the universe that we experience is comparable to a ``seashore'' for a larger ``ocean'' in which inaccessible processes take place. \footnote{The following quote is due to Isaac Newton  \cite{spence1858anecdotes}: ``I do not know what I may appear to the world, but to myself I seem to have been only like a boy playing on the seashore, and diverting myself in now and then finding a smoother pebble or a prettier shell than ordinary, whilst the great ocean of truth lay all undiscovered before me.''.}  
In this section we explore the conceptual consequences of this alternative picture.  We first briefly compare it to other interpretations of quantum mechanics, and then draw out some of the metaphysical implications.\footnote{Most of the implications were presented previously in \cite{thron2021sliced}.} Finally, we speculate about additional changes of viewpoint that could lead to a simpler and more comprehensive theory.

\subsection{Contrasts with other quantum theories}
There are a number of variant interpretations of QM, but the main ones are \cite{collins2007many}:
\begin{itemize}
\item
The Copenhagen Interpretation (and variants);
\item
The many-worlds interpretation;
\item
The pilot wave (de Broglie-Bohm) interpretation;
\item
The transactional interpretation;
\item
Consistent histories;
\item
Superdeterminism (a relatively recent contender \cite{hossenfelder2020rethinking}).
\end{itemize}

Besides consistent histories, these all presume causality (the transactional interpretation also has retrocausality). All view physical processes as involving influence traveling through time from causes to effects, either as a particle or wave packet. In some cases, this requires introducing a mechanism for collapsing the wave packet upon measurement. the extraverse requires no collapse, because our physical spacetime corresponds to a submanifold or ``slice'' of an ongoing process that does not terminate upon measurement.

\begin{figure}[h]
\begin{center}
\includegraphics[width=4.in]{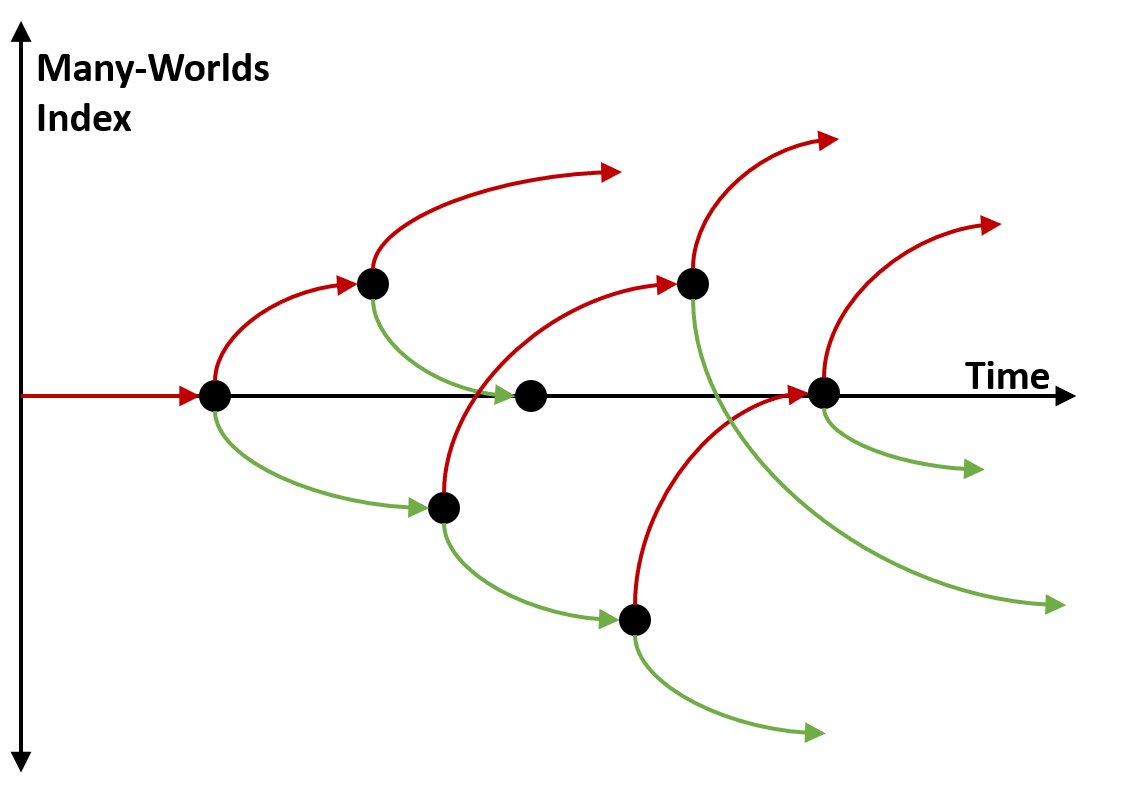}
\end{center}
\caption{ Many-worlds version of a 1-dimensional spacetime The branching is due to the exponential multiplication of alternative worldlines due to decoherence.(Compare the extraverse model in Figure~\ref{fig:0d}, where no branching is required.) \label{fig:MW}}
\end{figure}

The many-worlds picture avoids collapse by postulating a complicated spaghetti-like structure on physical reality, of which our own experience follows just one strand (see Figure~\ref{fig:MW}). The mess of decohering strands arises due to the insistence on imposing a linear ordering on causes and effects. This is like trying to embed two dimensions in one dimension, which leads unavoidably to bizarre complexities. The complexities disappear when the spacetime universe is embedded in a higher-dimensional space, and the causal process unfolds in the extra-spacetime dimension, as shown previously in Figure~\ref{fig:process}.

The extraverse interpretation is most similar to consistent histories. Consistent histories says that no past ``exists'' prior to and independent of measurement, and that the measurement essentially brings the past into being. The extraverse model similarly depicts past events as intrinsically intertwined with measurement, and represents both as springing from a common origin outside of spacetime.   Furthermore, unlike consistent histories the extraverse explains the origin of quantum indeterminacy, including the Born interpretation of the absolute square of the wavefunction as a probability density.

\subsection*{Metaphysical implications of the extraverse}
\begin{itemize}
\item[\#1:]
\emph{Spacetime is a cross-section, not a process.}
We typically think of physical reality as a process that develops over time. In virtually all equations in physics, chemistry, biology etc. that involve time, time is the independent variable and other quantities are functions of time (Einstein's equations in general relativity are the only exception that I can think of). 
Of course this is because we experience the universe as an unfolding process in time.  But this is a terrible reason! We also experience the earth as stationary, which led ancient scientists astray for centuries. 

The extraverse model envisions a ``block universe'' \cite{petkov2006there,vaccaro2018quantum}  in which time is no longer singled out as special, and instead is placed on the same level as dimensions of space. Time seems different to us only because of the way we experience the universe.  For example, if you're riding a train things appear to be moving in one direction. But that's just because we're riding the train. If we were riding a different train, things would look like they're moving in a different direction. In the same way, things seems to develop in time just because of how we humans are `travelling' through spacetime.

Unfortunately, the extraverse model doesn't show us how to predict the long-term future or access the distant past. It can't help us build a time machine or a faster-than-light transporter. We're stuck in our universe, and the picture doesn't show us how to jump off into the process dimension to see the processes developing. But on the other hand, it doesn't say that we \emph{can't} jump off!

\item[\#2:]
\emph{Entanglement is uncovered, not produced.}
Einstein's ``spooky action at a distance'' is a consequence of quantum entanglement, in which  particles or systems become mysteriously and invisibly ``linked'' due to  interactions. The extraverse model shows that interactions do not ``produce'' entanglement---rather, both the interaction and the measurements of entangled particles are vestiges of the same extra-universal process. 

\item[\#3:]
\emph{There are no causes without effects.} In the extraverse, what we typically call ``causes'' and ``effects'' share a common origin outside of the universe. When telescopes detect radiation from distant galaxies, the detection was not an ``afterthought'', but was intrinsically linked with the light's emission thousands of years ago. Similar ideas have been proposed before, by Wheeler and Feynman\cite{wheeler1945interaction} and earlier by Tetrode\cite{tetrode1922causal}.  
However, they did not envision the cause-effect connection as the culmination of a process that takes place outside of the universe.

\item[\#4:]
\emph{There is no ``reality'' apart from observation. }
We typically think of an experiment as an observer looking at some phenomenon or physical system from the outside, and making impartial, passive  observations and measurements. This concept was already roughed up by quantum mechanics, which showed that quantum measurements are never passive, and always disturb the system they measure. The extraverse model goes one step further, and proposes that this is because there is no objective reality ``out there'' to measure.  The division between ``phenomenon'' and ``measurement'' is somewhat arbitrary, since both are aspects of a larger whole. 


One may raise the objection that reality must exist independent of observation, since different observers always observe the same reality. However, this is a magic trick that is due to the common origin of all measurements within the extraverse. 
 
\item[\#5]\emph{Spacetime is not foaming.}
Conventional quantum field theory takes the view that (as Lawrence Krauss puts it) ``Empty space is a boiling, bubbling brew of virtual particles that pop in and out of existence in a time scale so short that you can't even measure them .'' \cite{KraussNothing}
The extraverse model interprets this supposed storm of activity  as a mathematical trace of the unfolding process located outside of accessible spacetime.

\item[\#6:] \emph{Free will and physics are reconcilable}
Free will is rejected by many scientists based on the premise  that past physical processes causally determine our present and future \cite{harris2012free}. But this apparent determinism is an illusion, says the extraverse model. Past, present, and future are not linked in an inexorable linear chain. 
There are underlying processes that determine your character, thoughts, and decisions, but they do not appear within this universe. 

One might object that the model simply displaces determinism into another dimension, and that I am still the outcome of processes outside of myself. But what is meant  by ``myself''?  In the extraverse model, the meaning of ``self'' is no longer so clear. How far do ``I'' extend into the process dimension? To say that ``I'' am confined to spacetime is like saying that an iceberg is only the visible part that sticks above the water, or a person is only the layer of skin that is visible.

\item[\#7:] \emph{The quest for a``Theory of Everything'' has no end.}
Consider the evolution of the conception of the physical world in Western intellectual history.
Greek philosophers conceived of solid objects as space-filling entities having extension and rigidity as fundamental properties. Much later, objects were re-envisioned as consisting of ensembles of jiggling atoms, held together by electromagnetic forces. Later, atoms themselves became  mostly empty space, consisting of much smaller particles (with electrons having no size at all). Then these particles turned into quanta, which are a strange amalgam of particle and wave. Later, quanta turned into resonances of a field, and different particles corresponded to different components of a multidimensional field.  The basic conceptual frameworks of each succeeding (supposedly) ``fundamental'' theory are radically different---these theories are not converging, but rather are wandering endlessly through an infinite wilderness of conceptual possibilities.
 
 By now, it should be abundantly clear that every physical theory is simply a convenient conceptualization of our current level of interaction with the universe. We are never getting any closer to ``true reality'' (which is a concept that should be banished from the scientific/philosophical vocabulary).  The best we can do is provide a series of analogies, supported by mathematical substructures. These are all temporary structures, and  will always be superseded. In a related context,  Ludwig Wittgenstein gave the analogy of a ladder, which is used to climb up to a higher level and is then discarded as unnecessary.
 
 \item[\#8:] \emph{Science explains almost nothing.}
Physical science is concerned with relationships between events, phenomena, and conditions in spacetime.  But the extraverse model shows that spacetime itself is only a single slice of extraverse, and, as such, has no direct access to the bulk of processes that rule the universe. As such, science can only offer explanations and predictions that are highly generalized or narrowly localized in space and time. Consider for example the famous ``baby picture of the universe'' obtained from the Hubble space telescope (see \cite{NASA}), which shows variations in the temperature of the microwave background radiation of the universe on the order of 200 microkelvins. Cosmic inflation theory magnificently predicts  the observed patchiness, but it cannot even begin to explain the location or extent of a single patch. At the small end of the scale, lattice quantum chromodynamics (QCD) simulations of a single hadron require computing power equivalent to 6,000 laptops operating continuously for a century \cite{berkeley2018}. How many universes filled with laptops would be required to simulate one milligram of material with more than $10^{20}$ hadrons (plus electrons)?

\item[\#9:]\emph{The physical universe is not a closed system.}
 Quantum ``randomness'' is due to information that is continually passed in from the encompassing extraverse. Furthermore, information from the physical universe is passed back out to the extraverse in the process dimension. 

\item[\#10:]\emph{Consciousness may be extra-physical}
 There are phenomena within the physical world which may be traced back to events outside spacetime, and thus cannot be reconstructed by  spacetime-bound procedures. Consciousness may be one such phenomenon. So we may not be able to build conscious computers.

 \item[\#11:]\emph{Spacetime is a selection of consciousness}
Our conscious experience flows along with one particular current of correlated events, which is situated within a much larger ``ocean'' (i.e. the extraverse).  There are myriads of other information flows within the extraverse, and our spacetime experience is just one particular selection.  This idea has both similarities and differences with the idea proposed (but apparently not entirely believed) by Wigner that consciousness causes the collapse of the wave packet during the measurement process\cite{ballentine2019meeting}.

\item[\#12:]\emph{There's room for angels.}
In the extraverse, there are other information flows besides that which our consciousnesses are capable of tracking.  This leaves the way open for other consciousnesses that possibly interact with our spacetime universe without actually belonging within it.

\end{itemize}

\subsection{Is momentum space more fundamental than spacetime?}
In the introduction, we proposed that mathematical complication in physical theories is often due to confusing observational viewpoint with basic physics. (This is by no means a new idea, for Isaac Newton said the same thing: ``Truth is ever to be found in simplicity, \& not in the multiplicity \& confusion of things. As the world, which to the naked eye exhibits the greatest variety of objects, appears very simple in its internal constitution when surveyed by a philosophic understanding, \& so much the simpler by how much the better it is understood, so it is in these visions.''\cite{newtonApoc}.)
In the above discussion we haven't really dealt with the issue of mathematical complication, because we began with the path integral formulation which is  fraught with mathematical difficulties when characterized rigorously \cite{albeverio2020rigorous}. Furthermore, there is a great deal left unexplained by the path integral formulation. It does not explain the quark or lepton masses, which are free parameters; neither does it explain why these ``particles'' obey Fermi statistics (e.g. the Pauli exclusion principle).  
Indeed, if fermions are considered as moving particles that can assume different momenta, then Fermi statistics are highly mysterious (what then would prevent two fermions from having the same momenta?). However, they are quite natural if we identify the appearance (or non-appearance) of a ``particle'' of fixed momenta as the inclusion (or exclusion) within spacetime of a latent phenomenon that exists within the larger extraverse.

There are several other indications that momentum space (rather than spacetime) affords a simpler and more revealing picture than spacetime. Field quantization is much more easily accomplished in momentum space rather than spacetime (\cite{wilcox2016macroscopic}, Section 14.2). In spacetime, field operators have support at vanishingly small points; while in momentum space the creation and annihilation operators correspond to waves that extend throughout space, which is much easier to represent mathematically and conceptually. 

Both the Schr{\"o}dinger and Dirac equations were motivated by dispersion relations expressed in momentum space. Dirac's dispersion relation comes from Einstein's famous energy-momentum relation
\begin{equation}
    (mc^2)^2 = E^2 - c^2|\vec{p}|^2
\end{equation}
and making the identifications $\vec{p}\rightarrow \hbar \vec{k}$ and $E \rightarrow \hbar \omega$ (due to de Broglie and Einstein, respectively), where $\vec{k}:= (k_1,k_2,k_3)$ and $\omega$ are wave number vector and angular frequency, respectively. In natural units $(c = \hbar = 1)$, we end up with:
\begin{equation}
    m^2 = \omega^2 - |\vec{k}|^2 \quad \textrm{or} \quad m^2 + k_1^2 + k_2^2 + k_3^2 = \omega^2.
\end{equation}
The wave numbers $k_1, k_2, k_3$ can be considered as ``freqencies'' (in space, rather than in time), this suggests that mass also corresponds to a frequency, possibly in the extra dimension that is posited by the extraverse model. The simply Pythogorean expression for frequency  also suggests that the distinction between timelike and spacelike dimensions is an observational artifact.

In his lecture ``The End of Spacetime'', Nima Arkani-Hamid  has pointed out that some transition amplitude calculations are far easier in momentum-energy space than in spacetime (the spacetime calculations require computing multitudes of terms that end up cancelling)\cite{Nima}.  

These observations leave the cumulative impression that spacetime itself may be an observational artifact, and more fundamental and comprehensive explanations may possibly be obtained by looking at momentum space instead.

\section{Summary}
Our main conclusion was anticipated by Shakespeare over 400 years ago:
\begin{quote}
There are more things in heaven and earth, Horatio,
Than are dreamt of in our philosophy. (\emph{Hamlet}, Act 1 Scene V)
\end{quote}

\bibliographystyle{plain}
\bibliography{main}

\end{document}